\documentstyle[aps,prl]{revtex}
\begin{document}
\title
{
Erratum: Effective Lorentz Force due to Small-Angle Impurity
Scattering: \\
Magnetotransport in High-$T_c$ Superconductors \\ {}
[Phys. Rev. Lett. {\bf 86}, 4652 (2001)]
}

\author{C.M. Varma and Elihu Abrahams}

\maketitle
\vskip .2in

Although the equations in this Letter (cond-mat/0011020) are correct, V.\
Yakovenko
\cite{yak} has pointed out to us that our evaluation, in the first
paragraph of page 4655, of the new contribution to the Hall angle is
incorrect. This estimate was based on the assumptions that the
angle-dependent impurity scattering rate $1/\tau_u$ had a non-zero
derivative at the zone edge and that all other angular dependence in
Eqs. (21-23) can be neglected for simplicity. Actually, as pointed out
by Yakovenko, $1/\tau_u$ must have zero derivative at the zone edge,
and then our estimate of the new contribution to the Hall angle vanishes.

The effect we derived is non-zero in a the general case
where other angular dependences are kept explicitly, such as the
angular anisotropies of the density of states and of the small-angle
cutoff parameter $\theta_c$. In fact, these are the microscopic basis
for the anisotropy of $1/\tau_u$; they give a contribution of the same
order as that of $1/\tau_u$ and must be kept in a consistent
calculation. These issues and a revised estimate will be addressed
elsewhere.

Fig.\ 1 concerns Ni-doped YBCO. The text and caption are
in error in referring to Zn doping.

We thank V. Yakovenko for his interest and helpful correspondence.

\end{document}